\documentclass[12pt,prb,preprint,showpacs]{revtex4} 
\usepackage{graphicx}
\begin{document}
\preprint{Submitted to Applied Optics}
\title{ Forces on a nanoparticle in an optical trap}
\author{Kaushik Choudhury}
\affiliation{%
Department of Applied Physics,  \\ Shri G S Institute of Technology \& Science, Indore - 452 003 India
}
\author{Joseph Thomas Andrews}
\homepage{http://www.sgsits.ac.in}
\email{jtandrews@sgsits.ac.in}%
\affiliation{%
Department of Applied Physics, \\ Shri G S Institute of Technology \& Science, Indore - 452 003 India
}%
\author{Pranay Kumar Sen}
\affiliation{%
Department of Applied Physics, \\ Shri G S Institute of Technology \& Science, Indore - 452 003 India
}%
\author{Pratima Sen}
\affiliation{Laser Bhawan, School of Physics,  Devi Ahilya University, Khandwa Road,
Indore - 452 017 India 
}%
\begin{abstract}
Forces on a nanoparticle in an optical trap are analysed. Brownian motion
is found to be one of the major challenges
to trap a nanoparticle. Accordingly,  suitable spatial electric
field distribution of laser beam is suggested
 to enhance the trapping force on a nanoparticle. Theoretical analysis is
carried out  to obtain conditions
for stable optical trap by incorporating the temperature variation at
large laser intensities. Numerical analysis is
made for single quantum dot of CdS in buffer solution irradiated by
an Nd:YAG laser. 
\end{abstract}

\maketitle

\section{Introduction}

Single laser beam optical tweezers are widely employed for trapping and manipulation
of micrometer sized particles. It finds variety of novel applications including  biomedical measurements \cite{bio}, quantum computation \cite{babcock}, single molecule nonlinear optical measurements~\cite{nlo},  etc. Of late, availability of sophisticated
instruments made it possible to realise optical manipulation of particles
with force and position  resolutions  of 0.1 pN and 0.1 nm \cite{stable}. With appropriate design and usage of modern electro-optic devices
such as spatial light modulator (SLM), piezo stages and high resolution charge coupled devices (CCD), manipulation of optical beam with custom designed spatial profile is possible \cite{grier,dufresne,bukusoglu,zhao}. 

Grier and his group \cite{grier} have used computer-generated holograms with SLM to create arbitrary three-dimensional configurations of single-beam optical traps that are useful for capturing, moving, and transforming mesoscopic objects in
3-dimensional holographic optical tweezers (3D-HOT). Through a combination of beam-splitting, mode-forming and adaptive wavefront correction,  HOT can exert precisely specified and characterized forces and torques on objects ranging in size from a few nanometers to hundreds of micrometers.
Dufresne et al~\cite{dufresne} created arbitrary configuration of optical tweezers using computer
generated diffractive optical elements with appropriate design to eliminate
stray lights. Recently, interest has been shown to construct optical traps with custom designed laser profiles \cite{bukusoglu,zhao,hoogenboom,svobo,
sugi,parkin,schnelle,macdonald,nano}. Optical tweezers with a haptic device and XYZ piezo scanner to manipulate
the position of microspheres are also used \cite{bukusoglu}. Zhao et al  \cite{zhao} used highly focussed hollow Gaussian beams to trap and manipulate Rayleigh particles. Hoogenboom
et al\cite{hoogenboom} have developed a method to create a pattern of wide variety of colloidal particles of size as small as 79nm with a resolution below the particle size using optical tweezers.  Today, one of the major interests in the field of optical trapping has been to manipulate smaller nanoscale
objects. For metallic substances, the trapping of a single 36-nm diameter gold particle
has been achieved \cite{svobo} and a 40-nm diameter gold particle was employed as a probe of scanning near-field optical microscope (SNOM) \cite{sugi}.
There are many reports which make use of arbitrary
custom designed laser profiles such as  non-circular Gaussian LG$_{02}$ mode \cite{parkin},
torroidal shape trap \cite{schnelle}   rotating interference pattern \cite{macdonald} establishing the usefulness of the arrangement to
trap transparent dielectric nano-particles successfully.

The major challenges
encountered in trapping nano-particle include (i) Brownian motion arising
due to thermal force and (ii) cluster formation. In dual beam optical tweezers
two types of temperature control methods were proposed by Mao et al \cite{thermal}.
They could overcome  the problem of heat convection at high intensities of
laser using a circulation method. On the other hand cluster formation occurs
when many particles are trapped simultaneously giving rise to larger finding
binding forces \cite{nano}. In our opinion, a detailed theoretical
understanding on nano-particle trapping with  optical
tweezers is still in its infancy and warrants attention. 
Present paper,
makes an attempt to understand the problems associated with trapping a nano-particle and suggests a suitable spatial profile of laser to enhance the trapping force. We have examines trapping conditions for various spatial profiles
of lasers viz., Hermite Gaussian (HG), Laguerre Gaussian (LG) and Custom designed
spatial profiles, giving due considerations to gradient, scattering
and thermal forces. A\ comparison of the results has been made which suggests
the suitability of custom designed modes over the HG and LG modes.

\section{Theory}

Optical tweezers take advantage of radiation pressure exerted on a micro-nano
particle to
trap it in an electromagnetic potential well. Radiation pressure is generated
by tightly focussing laser beam with a large numerical aperture (NA $\geq
1.0$) condensing lens or by using a combination of aspheric lenses. The radiation pressure  near the  focus arises due to (i)~scattering force
which is proportional to the intensity of light and is on the same direction
as that of the incident light and (ii)~gradient force which is proportional to the  spatial gradient of light intensity. 
For axial stability of a single beam trap, the ratio of backward gradient
force to forward scattering force should be greater than unity. As an additional
trapping condition for Rayleigh particle, we also need to take into account
the thermal forces. Apart from this precaution must be taken to avoid cluster
formation under suitable illumination conditions.

In the Rayleigh regime, when the particle size ($a$) is very small in comparison
to the wavelength ($a \gtrsim \lambda$), the scattering force is given by
\cite{ashkin}  
However when particle density is larger (or when mean free path is smaller)
optical binding force also becomes dominant.  The reported size of particle
trapped using conventional optical tweezers is 50nm to 100$\mu$m. The lower
limitation of size are attributed to the diffraction-limited trap volume,
smaller scattering and absorption cross sections and significant Brownian
motion due to large thermal energy acquired  by the particle. In order to
trap particle size much lower than the wavelength of the trap beam, the
challenges to be met are (i)~diffraction limited optics, (ii)~elimination
of optical binding forces and (iii)~overcoming the forces arising due to
Brownian motion of the nanoparticles. Accordingly, significant changes should
be made to the conventional optical tweezers design such as  holographic
optical tweezers \cite{grier}, haptic beam control \cite{bukusoglu}, electromagnic
coupled plasmonic optical trap \cite{grigo} or near resonant excitation
\cite{reso}.

Under Rayleigh scattering regime ($a\ll\lambda$), when the particle to be trapped is much smaller than the wavelength of the
light  used to generate radiation pressure, 
the various forces are calculated by assuming as
point dipole scatterers and the focus as a diffraction-limited region.  With this approximation the scattering and gradient forces are separated comfortably and can be analyzed independently. The scattering force for a point dipole
of size $a$ is found to be 
\begin{equation}
F_s=\frac{\sigma n_m}{c}I, \label{Fscat}
\end{equation}
where, $n_m$ is refractive index of the medium surrounding the nanoparticle,
$I$ is the intensity of light used to trap the particle and $\sigma$ is the
scattering cross section defined as 
\begin{equation}
\sigma = \frac{128\pi^5 a^6}{3\lambda^4}\left(\frac{n_r^2-1}{n_r^2+1}\right)^2,
\end{equation}
Here, $n_r$  is the relative refractive index defines as $n_r
= n_p/n_m$, with $n_p$ being the refractive index of the particle. The gradient force is given by \cite{ashkin}
\begin{equation}
F_g=\frac{2\pi  \wp}{cn_m^2}\nabla I \label{Fgrad}
\end{equation}
with the polarizability ($\wp$) of the particle being  
\begin{equation}
\wp = n_m^2 a^3\left(\frac{n_r^2-1}{n_r^2+1}\right).
\end{equation}
From eqs. (\ref{Fscat}) and (\ref{Fgrad}), it is evident that the scattering and gradient forces are proportional to $a^6$
and $a^3$, respectively. In other words, the magnitudes of both the scattering cross section and polarizability become too small to hold the particles inside
the trap. Apart from this, to trap a particle of dimension smaller than the wavelength of
light, one needs to take extra care of forces arising due to binding of particles
and thermal effects. The binding force dominates when the
particle mean free path is smaller. However, when
the particle density is smaller, the mean free path becomes larger and hence the binding
force may be ignored. With recent advances in piezo-stages, one can maintain a relatively stable equilibrium. For Rayleigh particles (of nanometer in size), thermal energy available
at room temperature is adequate  to dominate the scattering and gradient
forces and it becomes impossible to trap  the particle. In  his celebrated paper, Ashkin \cite{ashkin}  suggested that as an additional condition for trapping the Boltzmann factor ``$\exp(-U/k_BT)\ll1$'', with $U \;[\,=n_b \,\wp\, E^2/2]$ being the potential of the gradient force. Taking into account the thermal
conductivity, spot size and the optical absorption coefficient the corresponding
thermal force is found to be \cite{thermal}   
\begin{equation}
F_{Th}=\sqrt{2 k_B T \gamma_o}.
\end{equation}
where $\gamma_0 (=6\pi\rho \nu a)$ is the Stoke's friction constant with $\rho$
and $\nu$ as the mass density and kinematic viscosity of the medium, respectively.
For a nanoparticle, force arising due to the thermal energy will diffuse the Rayleigh particle. Accordingly, we consider that this force is acting isotropically in
all the directions.
Since a cw laser  is used for trapping particles a equilibrium in temperature
is expected to occur. The temperature of the medium / particle
is $T = T_0 + T'$, with $T_0$ is the ambient temperature while the increase
in temperature due to trap beam can be calculated from \cite{steen}
\begin{equation}
 T' = \frac{P_{in} (1-e^{-\alpha d})}{{2}{\pi}w_0\kappa}.
\end{equation}
Here, $P_{in} = P_{0} (1-R)$ with $P_0$ as the cw laser power, $R$ being the reflectivity of the medium while $\alpha$ and $\kappa$ are the  optical absorption coefficient and thermal conductivity of medium, respectively. $w_0$
and $d$ are spot radius at beam waist and thickness of the medium, respectively. 

For a given cw laser of power 100mW and a diffraction limited spot size of  50$\mu$m the force arising due to thermal energy
is independent of particle size and is calculated to be 0.4~pN at room temperature.
This magnitude of force is negligible as compared to the gradient and scattering
forces acting on a microparticle. On the other-hand, for a nanoparticle this is significant enough to dominate the
scattering and gradient forces. We introduce a dimensionless force parameter $F_D$ to
calculate the effective force on a particle by defining    
\begin{equation}
F_D=\left|\frac{F_g}{F_s + F_{Th}}\right|. \label{Ftot}
\end{equation}
Estimation of $F_D$  immediately gives the condition for the particle to be
trapped as $F_D \geq 1$ \cite{stablity}. From the above equation, it is clear that when the thermal force dominates,
the total force will be less than 1, leading to an unstable trap. In order to overcome this problem of trapping nanoparticles at room temperature,
we compare the radiative forces arising from some well known spatial profiles of laser beam
with suggested
custom designed laser beam profile which is possible
using suitable SLM and software code. Accordingly, the radiation pressure on a nanometer sized spherical particle is calculated by assuming three spatial profiles for the electromagnetic field, viz., (i) Hermite Gaussian (HG)
(ii) Laguerre Gaussian (LG) and (ii) a Custom designed
Gaussian profile (CG). We use standard equations under cylindrical coordinate
system to define the radiation profiles as \cite{parkin,stablity} 
\begin{eqnarray} 
u_{L}(r,\theta,z)&=&\frac{w_0}{w(z)}
\rho(z)^{l} L_{p}
^{l}[\rho(z)^2]\cos(l\theta)e^{-\rho(z)^2}
e^{-jkz-j\frac{kr^{2}}{2R(z)}-jl\theta},\\
u_H(r,\theta,z)&=&\frac{w_0}{w(z)}
H_p\left(\frac{\sqrt{2}r\cos\theta}{w(z)}\right)
H_l\left(\frac{\sqrt{2}r\sin\theta}{w(z)}\right)
e^{-\rho(z)^2}
e^{-jkz-j\frac{kr^2}{2R(z)}-jl\theta},\\
\hskip-5cm\hbox{ and}\hskip38mm&&  \nonumber \\
u_C(r,\theta,z)&=&\frac{w_0}{w(z)}\left(1-\cos\left[\sum_pe^{-(\frac{\theta-p\theta_0}{d\theta})^2}\right]\right)\sum_le^{-(\frac{r-lr_0}{w(z)})^2}
e^{-jkz-j\frac{kr^2}{2R(z)}-jl\theta}.
\end{eqnarray}\label{beam}
Here, $\rho(z)=\sqrt{2}r/w(z)$, $R(z) = z+iz_R$ is the radius of curvature
with $z_R\;(=\pi w_o^2/\lambda)$ being the depth of focus and spot size $w(z)
= w_{0}\sqrt{1+z/z_R}$. $p,\; l \;(= 0, \;1,\;2 \;...)$ are integers
defining laser modes. $H_{p\,(l)}$ is Hermite Gaussian function of order
$p\, (l)$. $L_p^l$ is the generalised Laguerre Gaussian of order $l$ and index $p$. The intensity of the laser beam may be calculated from
\begin{equation}
I_i(r,\theta,z) = I_0 |u_i(r,\theta,z)|^2, \label{int}
\end{equation}
with $i = L,\,H,\, C$, representing LG, HG or CG laser profiles, respectively. For $p=l=0$,
all profiles defined above reduce to TEM$_{00}$ mode structure as depicted
in Figure 1. The figure also shows other profiles obtained for $p = l =$ 3, 6  and 9. One can notice from the figure that for higher order modes the beam profile becomes sharper while the individual linewidths of each mode is not a constant. In other words, for higher order modes one can expect sharp rise to
the magnitudes of gradient and scattering forces
when the number of modes increase. We make use of this feature as
a key point to explore the possibility
of trapping nanoparticles in a single beam optical trap. 

\begin{figure}[htb]
\begin{center}
\includegraphics[width=14cm]{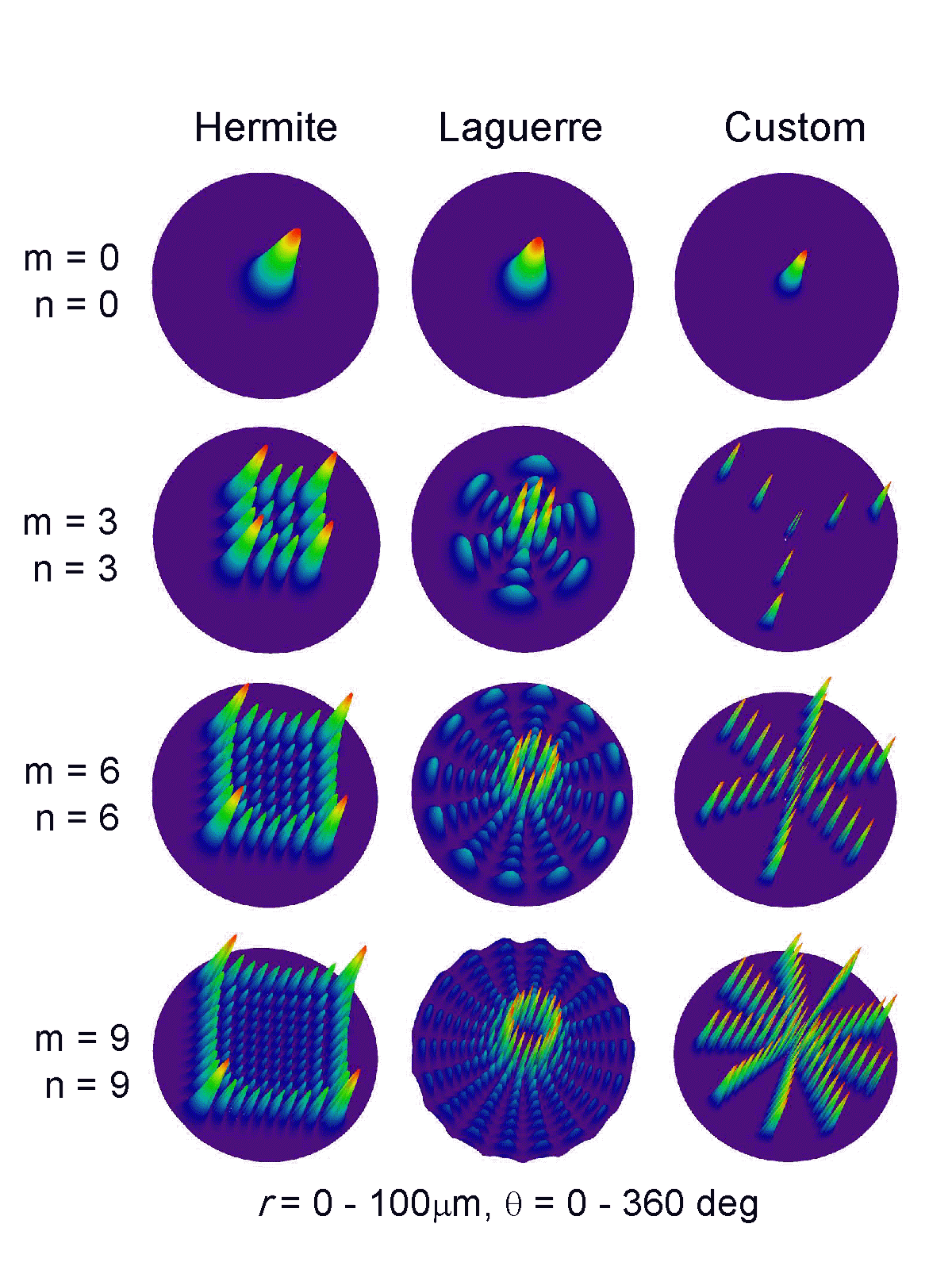}
\end{center}
\caption{Spatial intensity profiles of various beam geometries obtainable
from equation (\ref{beam}). }
\end{figure}

\section{Results and Discussions}
In order to understand the present set of equations for trapping a nanoparticle,
we have calculated the net force acting on a semiconductor single quantum dot (SQD) of CdS suspended in a colloid matrix. 
\begin{figure}[ht]
\begin{center}
\includegraphics[width=14cm]{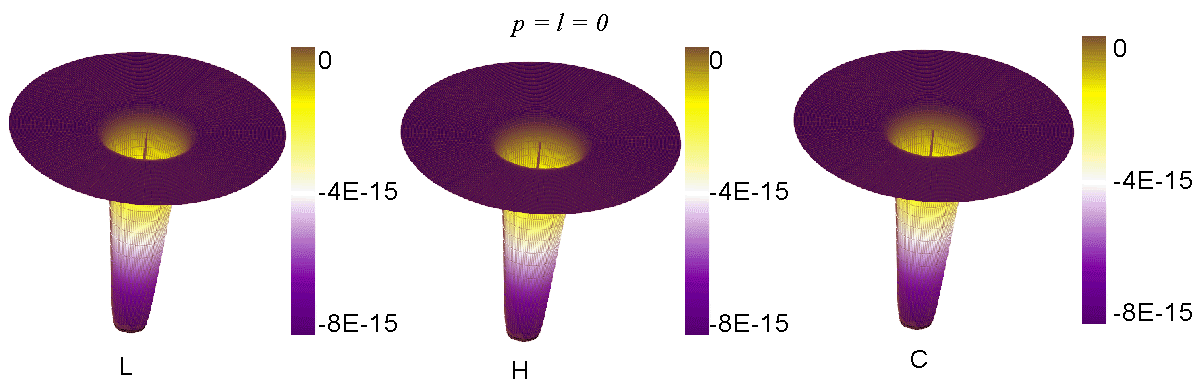}
\end{center}
\caption{Dimensionless force parameter $F_D$, plotted in cylindrical coordinate
system for  Laguerre~(L), Hermite (H) and Custom designed (CC)
 laser profiles as defined
in Fig. 1 for $p=l=0$.}
\end{figure}
The material parameters
are: $a$ = 5nm, $n_p$ = 1.65, $n_m$ = 1.33. $\kappa$ = 97 m/K, $\alpha$ = 15 $\times$ 10$^{-5}$cm$^{-1}$.
\begin{figure}[ht]
\begin{center}
\includegraphics[width=14cm]{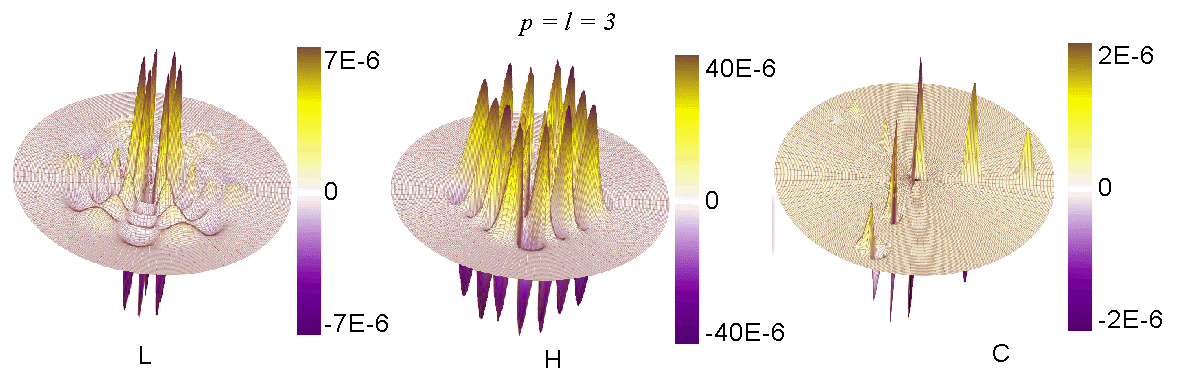}
\end{center}
\caption{Same as Fig. 2, but for $p=l=3$.}
\end{figure}
The laser assumed to be Nd:YAG or a diode pumped solid state laser operating
at 1.064 $\mu$m and having an optical power of 100mW. 
\begin{figure}[ht]
\begin{center}
\includegraphics[width=14cm]{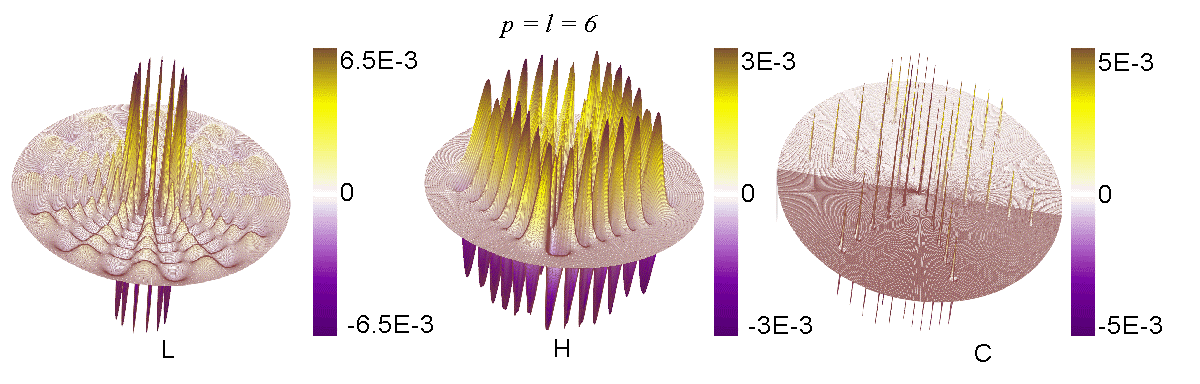}
\end{center}
\caption{Same as Fig. 2, but for $p=l=6$.}
\end{figure}

\begin{figure}[ht]
\begin{center}
\includegraphics[width=14cm]{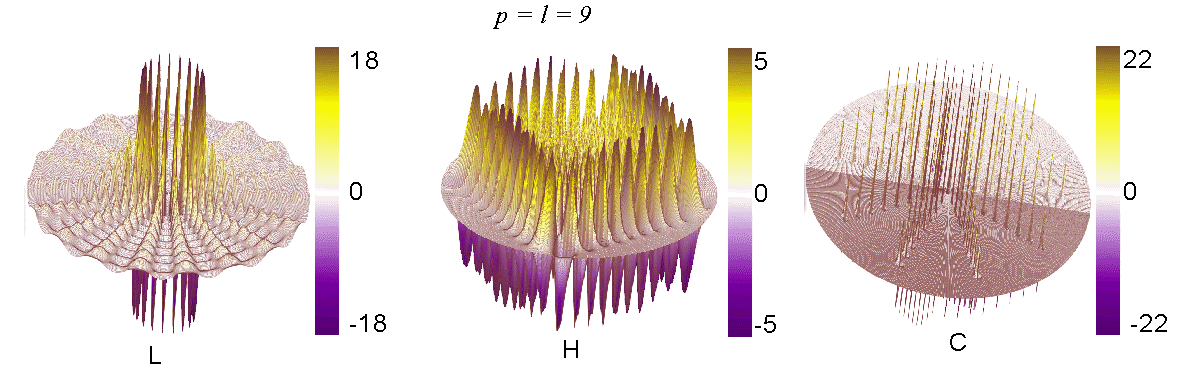}
\end{center}
\caption{Same as Fig. 2, but for $p=l=9$.}
\end{figure}

\begin{figure}[ht]
\begin{center}
\includegraphics[width=14cm]{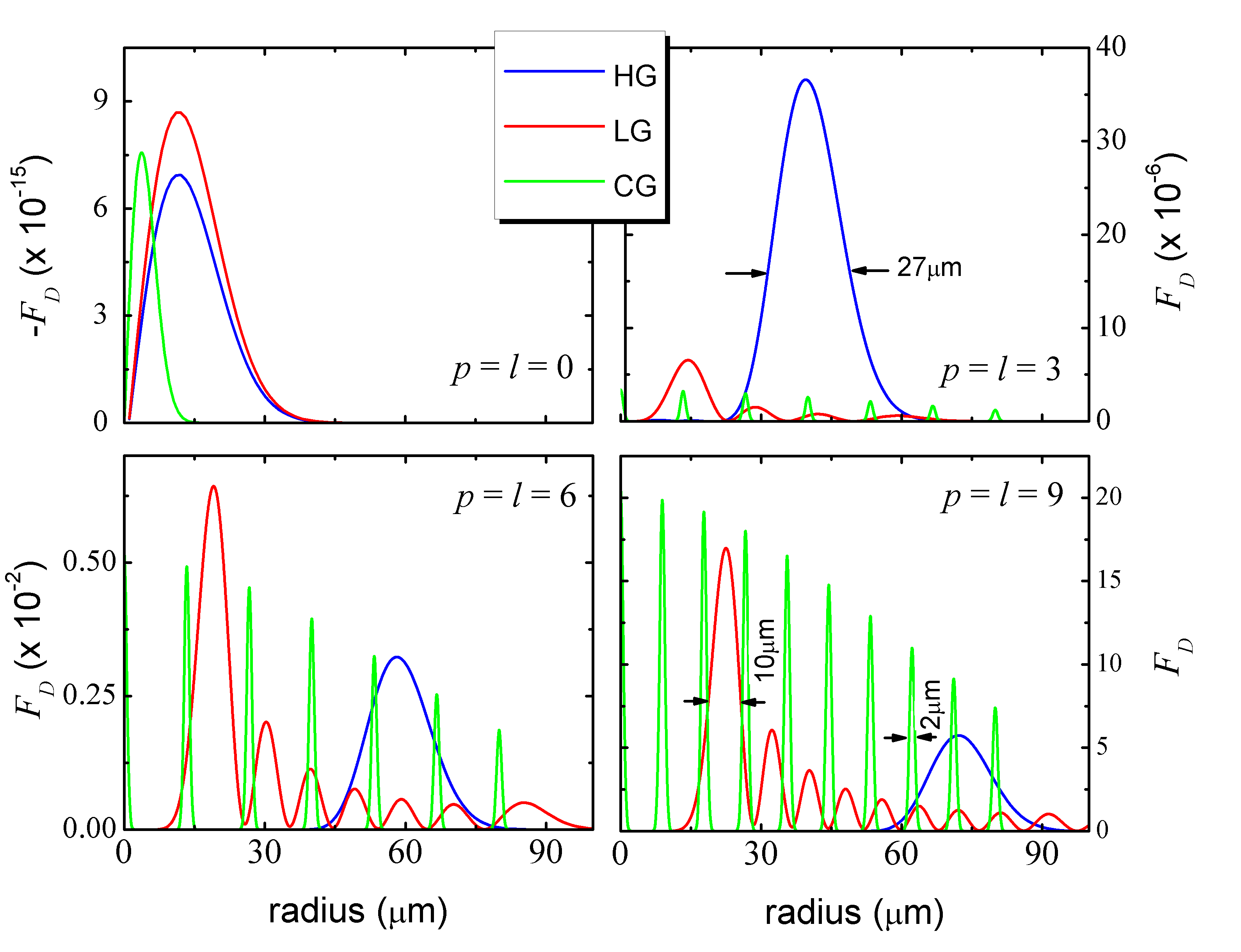}
\end{center}
\caption{Dimensionless force parameter  $F_D$ plotted in cylindrical coordinate
system for  Laguerre (LG), Hermite (HG) and Custom designed laser profiles
(CG) as defined in Fig. 1 for $p=l=0,\; 3, \,6$ and 9.}
\end{figure}

Equations (\ref{Ftot}) - (\ref{int}) are utilised in Fig.~2, to demonstrate the ratio of forces ($F_D$) acting on a transparent nanoparticle with a TEM$_{00}$ mode of laser beam  corresponding to  $p = l = 0$. On similar conditions of material parameters the total force
acting on the particle for $p = l = 3,\; 6 $ and $9$ are exhibited in Figures 3, 4 and 5, respectively. The magnitude of forces in these curves may be
attributed to the nature of laser beam profile as given in Figure 1. In these figures, we find that with increasing mode index from $p = l = 0$ to $p =
l = 9$, the ratio $F_D$ can be made to exceed the threshold condition required
(i.e. $F_D \geq 1$) to trap a nanoparticle in all three profiles. The nanoparticle
will be trapped within the boundary of the lobes obtained for various modes.

In order to obtain the area within which the particle can be trapped,we have
plotted $F_D$ as a function of radial distance from beam axis. From Figure
6, we could find that the nanoparticle can be trapped within circle of radius
approximately 13.4~$\mu$m, 5.1~$\mu$m and 1.1~$\mu$m, respectively for HG,
LG and CG beam profiles.
The custom designed profile therefore offers better trapping possibility
for a nanoparticle.
If particle density within the colloid matrix is such that only few nanoparticles
are available within that trap region, then a probability for achieving single
nanoparticle trap becomes possible using a custom designed laser profile.  If need arises, one can construct a custom designed laser profile
with appropriate numbers of modes of minimum
possible width to trap a particle. This could be done  with suitable program
code and addressing it to a high resolution spatial light modulator based on LCD or LCoS device \cite{hama}. 

\section{Conclusions}
To conclude, we proposed a method to trap submicron to nanometer  sized particles
in a single beam optical tweezers. The domination of external forces including thermal
force which could weaken the chance of trapping
nanoparticles are addressed properly. The proposed method suggests the usage
of  a properly addressed spatial light modulator combined with a nano-positioning piezo device to trap a nanoparticle inside an optical trap. The theoretical estimations for a single semiconductor quantum dots in a buffer solution confirms the possibility of trapping a nanoparticle in a single laser optical trap.   
\begin{acknowledgments}
The authors acknowledge the financial support received from UGC and DST,
New Delhi. KC thank BIT,
Mesra, Ranchi for leave.
\end{acknowledgments}


\begin{thebibliography}{99}
\bibitem{bio} A. Ashkin, J. M. Dziedzic, J.~E.~Bjorkholm and S.~Chu, {\em
Opt. Lett.}, 288-291, {\bf 11} (1986); S. B. Smith, Y. Cui and C. Bustamante,
{\em Science}, 795-799, {\bf 271} (1996).
\bibitem{babcock} N. S. Babcock, R. Stock, M. G. Raizen and B. C. Sanders,
{\em Can. J. Phys.}, 1-9, {\bf 99} (2008).
\bibitem{nlo} M. Gerlach, Yu. P. Rakovich, J.~F.~Donegan, N.~Gaponik, A.~L.~Rogach,
{\em Phys. Status Solidi C}, {\bf 3}, 3689-3692 (2006); {\em ibid}, {\em
Optics Express}, {\bf 15}, 3597-3606 (2007).
\bibitem{stable} K. J. Van Vliet, G. Bao and S. Suresh {\em Acta Materialia}, {\bf 51} 5881-5905 (2003). \url{http://www.lastek.com.au} Model - Nano-HS series., \url{http://www.pi.ws},
\url{http://www.physikinstrumente.com} Model - D-100.

\bibitem{grier} E. R. Dufresne and D. G. Grier, {\em Rev. Sci. Instr.}, {\bf
69}, 1974-1977 (1998);  D. G. Grier and Y. Roichman, {\em Appl. Opt.}, {\bf 45}, 880-887 (2006).
\bibitem{dufresne} E. R. Dufresne, G. C. Spalding, M. T. Dearing, S. A.
Sheets and D. G. Grier, Compuet Generated holographic  tweezers.
\bibitem{bukusoglu} C. Basdogan, A. Kiraz, I. Bukusoglu, A. Varol, and S. Do\~{g}anay, {\em
Optics Express}, {\bf 15}, 11616-11621 (2007).
\bibitem{zhao} C.-L. Zhao, L.-G. Wang, X.-H. Lu, {\em Phys. Lett. A}, {\bf
363}, 502-506 (2007).
\bibitem{hoogenboom} J. P. Hoogenboom, D. L. J. Vossen, C. F-Moskalenko,
M. Dogterom and A. van Blaaderen, {\em Appl. Phys. Lett.}, {\bf 80}, 4828-4830
(2002).
\bibitem{svobo} K. Svoboda, S. M. Block, {\em Opt. Lett.}, {\bf 19}, 930 (1994).
\bibitem{sugi} T. Sugiura, T. Okada, Y. Inoue, O. Nakamura, S. Kawata, {\em
Opt. Lett.}, {\bf 22}, 1663 (1997).
\bibitem{parkin} S. J. Parkin, T. A. Nieminen, N. R. Heckenberg and H. R-Dunlop,
{\em Phys. Rev. A}, {\bf 70}, 023816 (2004).
\bibitem{schnelle} S. K. Schnelle, E. D. van Ooijen, M. J. Davis, N. R. Heckenberg
and H. R-Dunlop, versatile two-dimensional potential 
\bibitem{macdonald} M. P. MacDonald, K. V-Sepulveda, L. Paterson, J. Arlt,
W. Sibbett and K. Dholakia
\bibitem{nano} Y. R. P. Sortais, H.~Marion, C.~Tuchendler, A.~M.~Lance, M.~Lamarc,
P.~Fournet, C.~Armellin, R.~Mercier, G.~Messin, A.~Browaeys, and P.~Grangier,
Diffraction limited optics for single atom manipulation; A. S. Zeleninia,
R. Quidant and M.N-Vesperinas, {\em Opt. Lett.}, {\bf 32}, 1156-1158, (2007).
\bibitem{grigo} A. N. Grigorenko, N. W. Roberts, M. R. Dickinson  and  Y. Zhang, {\em Nature Photonics}, {\bf 2}, 365 - 370 (2008). 
\bibitem{reso} T. Iida, H. Ishihara, {\em Phys. Rev. Lett.}, {\bf 97}, 117402 (2006); {ibid}, {\em Nanotechnology}, {\bf 18}, 084018 (2007).
\bibitem{ashkin} A. Ashkin, J. M. Dziedzic, J. E. Bjorkholm and S. Chu, {\em
Opt. Lett.}, {\bf 11}, 288-290 (1986).
\bibitem{thermal}  D. Selmeczi, S. F. T.-Nørrelykkec, E. Schäfferd,
P. H. Hagedorne, S. Moslera, K. B.-Sørensenf, N.~B.~Larsena,
H.~Flyvbjerg, {\em Acta Physica Polonica B}, {\bf 38}, 2407-2431 (2007); U.  F. Keyser, D. Krapf, B. N. Koeleman, R. M. M. Smeets,
N. H. Dekker, and C. Dekker, {\em Nano Letters}, {\bf 5}, 2253-2256 (2005).
\bibitem{steen} Steen H. Mao, J. R. A.-Gonzalez, S. B. Smith, I. Tinoco Jr.,
and C. Bustamante, {\em Biophysical Journal\/}, {\bf 89}, 1308-1316 (2005).
\bibitem{stablity} Y. Harada and T. Asakura, {\em Opt. Commun.}, {\bf 124},
529-541 (1996).
\bibitem{hama} \url{http://www.hamamatsu.co.jp, http://www.holoeye.de}
\end{thebibliography}
\end{document}